# Studying habitability of the exoplanents Kepler-504 b, Kepler-315 b, and Kepler-315 c

Sattik Bhaumik[1], Geetanjali Sethi[2]
[1] Don Bosco School, Siliguri, West Bengal, India
[2] St. Stephen's College, University of Delhi, Delhi, India

**SUMMARY**
**Astronomers have always wanted to know whether there are other planets around other stars that support life like our Earth. The search for life elsewhere has led us to new findings of extreme planetary conditions that humans were unaware of. Here we present the habitability index values of three earlier discovered exoplanets: Kepler-504 b (of the star Kepler-504), Kepler-315 b and Kepler-315 c (of the Kepler-315 stellar system). We wanted to know what are the ideal factors that decide the chances of habitability, e.g., the orbital distance from the star, the type of star, or a combination of multiple properties. We hypothesized that there should exist a combination of these properties that will increase the chances that the exoplanet would be similar to Earth. We have adopted the Earth Similarity Index (ESI) for calculating the physical similarity of exoplanets to Earth, and hence the probability of them being habitable. Using available telescope data, we found that Kepler-504 b, with a host M-type star (small red dwarf), and Kepler-315 c, with a host G-type star, had ESI values of 71.23% and 69.44%, respectively, thereby showing high similarity to Earth. Kepler-315 b, with a host G-type star, on the other hand, had an ESI value of only 35.68%, showing poor similarity to Earth. We have also listed previously calculated ESI values of additional exoplanets from the Planet Habitability Laboratory catalog, which supports our hypothesis. Thus, it suggests that a combination of star-type and orbital radius seems to make conditions favorable. Future work can study more exoplanets with such combinations to further validate these findings.**

**INTRODUCTION**
The idea of planets around other stars has been hypothesized since the Greek Civilization (1). Are there more solar systems just like ours? How many support life? These questions have motivated us to look for evidence of planets beyond our solar system, i.e., extrasolar planets or simply exoplanets, in the last few decades. The first exoplanets were discovered in 1992 by analyzing periodic Doppler shifts, which are the relative change in wave frequency in the stellar spectra with respect to an observer (1). Those spectral wave frequency changes revealed the two rocky planets orbiting the pulsar PSR B1 257+12, a rotating neutron star that emits electromagnetic radiation at regular intervals in the Virgo constellation (1). However, these planets were uninhabitable for organic life due to the continuous intense radiation of the pulsar bombarding them (1). The new techniques of Doppler shift and Radial Velocity of stars were, however, constrained by uncertainties and lack of precision (1, 2). This required longer periods of observation to determine the planetary parameters (radius, mass, etc.) (2). Astronomers have made rapid progress in telescopes since then (3).

Missions like Kepler Space Telescope (KST) and Transiting Exoplanet Survey Satellite (TESS) have revolutionized our way of thinking about planetary systems (3). In fact, they have questioned our theories of planetary formations and helped us refine them over the years. KST has been tasked with discovering Earth-like planets in habitable zones, e.g., regions in which water can exist in a liquid state on the planet's surface given an appropriate atmosphere, around main-sequence stars (4). KST has achieved breakthroughs in finding habitable exoplanets, new planetary systems that demanded completely new planet formation theories, and extreme exoplanets in terms of orbital radius, temperature, and atmospheric pressure (4). Similarly, TESS has been recently looking into a field 400 times larger than that of KST (2, 4). As the name suggests, it monitors periodic brightness dimming due to planet transits. These telescopes have had extended missions such as K2 for KST (4). KST has discovered the most exoplanets, more than 2600 to date. Astronomers are still analyzing the data KST has produced and discovering new planets (4). TESS has additionally gathered over 2100 probable candidates and will continue to observe more stars and provide stellar data in the coming years (4).

However, even with current techniques, finding Earth-like planets is difficult. Our Sun is an example of a G dwarf star, which has surface temperatures in the 4,726.85 - 5,726.85°C range but makes up only 3% of the stars in the universe (5). In order to find planets suitable for life, we need to look for planets at a certain orbital radius from other types of stars, closer to cooler stars and distant from hotter stars, due to the temperature suitable for carbon-based life (5). More common are M dwarfs, small red stars with a surface temperature <3226.85°C that make up 75% of all known stars (6). The second most common after M dwarfs are the K-type dwarf stars with a surface temperature 3226.85 - 4,726.85°C (6). Besides, there are the yellow/white F-type stars with temperatures ranging 5,726.85 - 6726.85°C (6). Also, there are exoplanets that have surfaces similar to Earth, i.e., rocky, and some have water in the form of ice (7). Rocky and oceanic surfaces are ideal for life to exist (7). Additionally, there are exoplanets that are entirely oceanic, with no solid rocky surface (7). The chances of complex life thriving there is good, but lower than those exoplanets with solid surfaces (6, 7).

We wanted to know what the ideal properties are, e.g., distance from a star, type of star, surface properties like Earth, or a combination of properties, that decide the chances of habitability. We hypothesized that there should exist a combination of these properties that will increase the chances that the exoplanet would be similar to Earth. We have adopted the Earth Similarity Index (ESI) for calculating the physical similarity of exoplanets to Earth, and hence the probability of them being habitable (8). We decided to estimate the ESI as a measure of habitability for exoplanets based on the kind of host star, and their orbital radius. We have calculated the ESI for four parameters (radius, density, escape velocity, and surface temperature) (8).

We hypothesized that there should exist a combination of these properties that will increase the chances that the exoplanet would be similar to Earth. Our results suggested a pattern, because we could see that Kepler-315 c, with a host G-type star and orbital radius very similar to Earth had an ESI greater than 60% and Kepler-504 b, with an M-type host star (small red-dwarf), and a smaller orbital radius than Earth also had an ESI greater than 60% (9, 10). Since M-type stars have comparatively lower temperatures, planets being closer to it ensure higher ESI. Kepler-315 b being closer to its host star makes it hotter and unsuitable for life (10). We also compared previously calculated ESI data for additional exoplanets taken from the Planet Habitability Laboratory (PHL), a database of all habitable planets, that showed support for our hypothesis (11). Thus, our results suggest that a combination of star type and orbital radius result in conditions favorable for life.

## RESULTS

We hypothesized that there should exist a combination of these properties that will increase the chances that the exoplanet would be similar to Earth. We have adopted the Earth Similarity Index (ESI) for calculating the physical similarity of exoplanets to Earth, and hence the probability of them being habitable (8). In order to study the parameter combination of habitable exoplanets, we analyzed data provided by the MAST portal with regard to radius, density, escape velocity, and surface temperature of the exoplanet and its star (12). The ESI values of four parameters (radius, density, escape velocity and surface temperature) were calculated. Then, we calculated Interior ESI ($ESI_i$) and Surface ESI ($ESI_s$) using those four parameters (radius, density for $ESI_i$ and escape velocity, temperature for $ESI_s$, respectively) (8). The habitability of the exoplanets was determined by calculating the Global ESI ($ESI_G$) (8).

The period of Kepler-315 b is around 96 days, as can be seen from the BLS power vs time data as the peaks in power are separated by roughly 96 days (**Figure 1A-B**). Using this, a folded light curve was generated which fits the observed data perfectly (**Figure 1C**). From the folded light curve, which was produced from this orbital period, we determined the orbital radius of the planet to be 0.402 AU (Astronomical Units, 1 AU = 149.6 million km) and its radius to be 5.22 times that of

|  | Earth | Kepler-315 b | Kepler-315 c | Kepler-504 b |
|---|---|---|---|---|
| F_trans value |  | 0.9979 | 0.9978 | 0.9992 |
| Planet Orbital Period (days) | 365 | 96.1 | 265.5 | 9.5 |
| Planet Orbital Radius (m) | 1.5e+11 | 5.6548e+10 | 1.11325e+11 | 9.10615e+9 |
| Planet Orbital Radius (AU) | 1 | 0.402 | 0.744 | 0.065 |
| Planet Radius (m) | 6.3781e+6 | 3.3296e+7 | 1.9602e+7 | 1.0384e+7 |
| Planet Radius (compared to Earth) | 1 | 5.22 | 3.07 | 1.628 |
| Planet Radius (compared to Jupiter) | 0.08921 | 0.466 | 0.27 | 0.145 |
| Planet Temperature (Celsius) | 15 | 182.1 | 51.31 | 114.84 |
| Planet Mass (kg) | 5.972e+24 | 8.182e+25 | 9.615e+25 | 1.881e+25 |
| Planet Mass (compared to Jupiter) | 0.00315 | 0.043 | 0.05 | 0.01 |
| Planet Mass (compared to Earth) | 1 | 13.7 | 16.1 | 3.15 |
| Planet Density (g cm$^{-3}$) | 5.52 | 0.53 | 3.05 | 4.01 |
| Escape Velocity of Planet (km/s) | 11.2 | 18.1 | 25.59 | 15.55 |

**Table 1. Exoplanet Parameters Dataset.** The parameters of all Kepler-315 b, Kepler-315 c, and Kepler-504 b calculated by the Python code along with host star type. Earth's data has also been included for comparison.

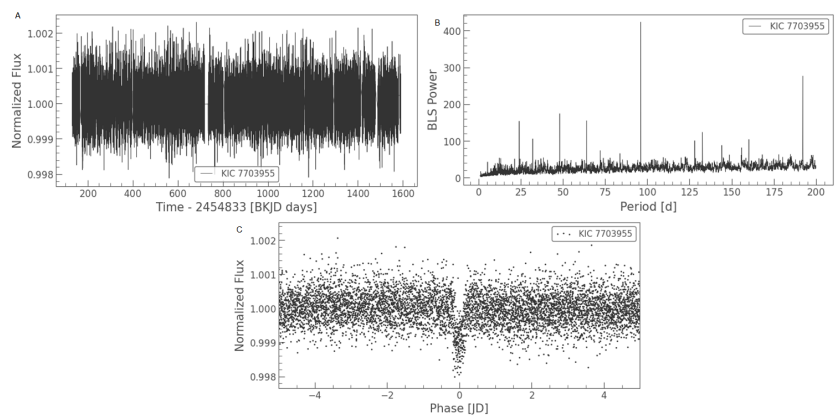

**Figure 1. Kepler-315 b flux, orbital period, and transit flux.** The prominent flux drop while observing the star Kepler-315 shows the close orbit of Kepler-315 b. **(A)** Normalized brightness (flux) of the star Kepler-315 over its whole observed time period. **(B)** Orbital period of the planet using the BLS method. The BLS power (transit signal, unitless periodogram object), of folded light curve is plotted over a period of days [d]. **(C)** The Transit Flux (Flux versus Phase) reveals the minimum flux value of planet Kepler-315 b, which helped us calculate its planetary properties and habitability.

Earth's radius (**Table 1**).

We first masked the Kepler-315 b data and repeated the calculations for Kepler-315 c and found the period of the planet to be around 265.5 days as per the BLS power vs time data (**Figure 2A-B**). Using this, the folded light curve was generated (**Figure 2C**). From this folded light curve, we determined the planet's orbital radius to be 0.744 AU and its radius to be 3.07 times that of Earth's (**Table 1**).

Similarly, we plotted the normalized flux of the Kepler-504 over its entire period of observation (**Figure 3A**). Applying the BLS periodogram method, we plotted the BLS power vs period data (**Figure 3B**). From the graph, we got the orbital period of Kepler-504 b to be around 9.5 days. Using this, we plotted the folded light curve and calculated the orbital radius of the planet to be 0.065 AU (**Figure 3C**). Then, we found the planet's radius to be 1.628 times that of Earth (**Table 1**).

Kepler-315 b has a surface temperature of about 182°C, Kepler-315 c has a surface temperature of about 51°C, and Kepler-504 b planet's surface is about 114°C (9, 10, **Table 1**). Since M-type stars have comparatively lower temperatures (<3226.85°C), planets being closer to it ensure higher habitable conditions (6). Kepler-315 b is at 0.402 AU while Kepler-315 c is at 0.744 AU from their host star (9). Meanwhile, Kepler-504 b is at 0.065 AU from its host star and has the highest density (4.01 g•cm$^{-3}$) and is the most similar to Earth's density among the three candidates (**Table 1**) (10).

Kepler-504 b and Kepler-315 c show high similarity of the $ESI_i$ (84.1% and 68.1%, respectively) and $ESI_s$ (60.11% and 70.81%, respectively) to that of Earth, while Kepler-315 b is not similar in either parameter to Earth (only 28.5% $ESI_i$ and 44.68% $ESI_s$ respectively) (**Table 2**). Kepler-504 b and Kepler-315 c are habitable with a probability of about 71.23% and 69.44% $ESI_G$, respectively, while Kepler-315 b with 35.68% $ESI_G$ appears unfit for organic-based life (**Table 2**).

We have included the data of seven exoplanets from five host stars Kepler-1680 (Kepler-1680 b), Kepler-520 (Kepler-520 b, Kepler-520 c), Kepler-1649 (Kepler-1649 b, Kepler-1649 c), Kepler-22 (Kepler-22 b) and Kepler-452 (Kepler-452 b) (13-15, **Table 2**). Among the seven additional exoplanets we examined from previous work, Kepler-22 b (71% $ESI_G$), Kepler-452 b (83.3% $ESI_G$), and Kepler-1649 c (92% $ESI_G$) showed the most similarity to Earth based on $ESI_G$ (**Table 2**). The Kepler-1649 system is similar to the Kepler-315 system with respect to our hypothesis. Kepler-520 and Kepler-1680 systems having an F-type and a K-type star respectively also show the need for a G-type or M-type to support life. But the analysis is not sufficient to establish any clear relationship between orbital radius, host star type, and ESI.

| Planet/Exoplanet | Host Star Type | Orbital Radius (AU) | Interior ESI | Surface ESI | Global ESI |
|---|---|---|---|---|---|
| Earth | G-type dwarf | 1 | 100% | 100% | 100% |
| Kepler-315 b | G-type | 0.402 | 28.5% | 44.68% | 35.68% |
| Kepler-315 c | G-type | 0.744 | 68.1% | 70.81% | 69.44% |
| Kepler-504 b | M-type | 0.065 | 84.1% | 60.11% | 71.23% |
| Kepler-22 b | G-type | 0.849 | 72.64% | 69.40% | 71% |
| Kepler-452 b | G-type | 1.046 | 80.56% | 86.12% | 83.3% |
| Kepler-1649 b | M-type | 0.051 | 25.52% | 73.45% | 43.30% |
| Kepler-1649 c | M-type | 0.064 | 89.11% | 94.89% | 92% |
| Kepler-1680 b | K-type | 0.078 | 5.66% | 19.53% | 10.51% |
| Kepler-520 b | F-type | 0.144 | 91.54% | 16.25% | 38.57% |
| Kepler-520 c | F-type | 0.059 | 15.60% | 6.05% | 9.72% |

**Table 2. Host Star Type, Orbital Radius and ESI (Interior, Surface and Global) values.** In addition to the three planets explored in this study and earth, data for planets Kepler-1680 b, Kepler-520 b, Kepler-520 cm Kepler-1649 b, Kepler-1649 c, Kepler-22 b and Kepler-452 b have also been included for comparison of planets with different host-star types and orbital radii (2, 10, 14-16). Planets with a host-star of M-type or G-type show a higher ESI value than other types. Orbital radius is another important factor as Kepler-315 b being nearer to its host star has lower ESI than Kepler-315 c, which is further away. The Kepler-1649 system also shows evidence for the importance of orbital radius.

## DISCUSSION

Our results suggest that Kepler-504 b and Kepler-315 c are habitable (71.23% and 69.44% ESI, respectively), while Kepler-315 b is not fit for organic-based life (only 35.68% ESI). These results support our hypothesis that there are viable combinations of host star type and planet orbital radius that increase the probability of a planet being habitable. This combination hypothesis also finds further support from the evidence of the additional seven exoplanets in the relation between their orbital radii, host star type and their $ESI_G$ (**Table 2**). According to the Planet Habitability Laboratory (PHL), an ESI value of 60% and above is considered habitable (11). The higher the ESI value above 60%, greater is its habitability to support Earth-based life (11). The highest ESI of a non-Earth planet, according to PHL, is that of HD 216520 c, Neptune-like, with an ESI of 63% (11).

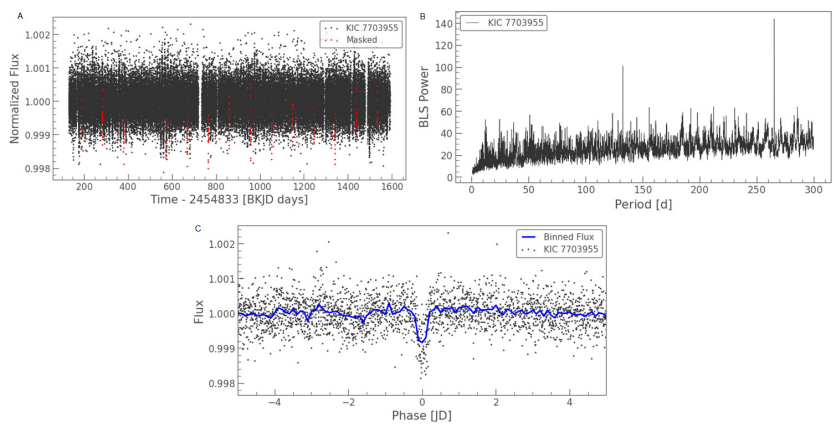

**Figure 2. Kepler-315 c flux, orbital period, and transit flux.** The slight flux drop while observing the star Kepler-315 shows the far away orbit of Kepler-315 c around it. **(A)** Brightness (flux) of the star over its whole observed time period. It is the same as Figure 1A except that Kepler-315 b data was masked (hidden) to calculate the orbital period for Kepler-315 c. **(B)** Orbital period of the planet using the BLS method. **(C)** The Transit Flux graph gives us the minimum flux value of planet Kepler-315 c which helps us calculate planetary properties and habitability. Flux was binned to show the transit curve we were looking for in case the curve was not prominent enough due to masking of data points.

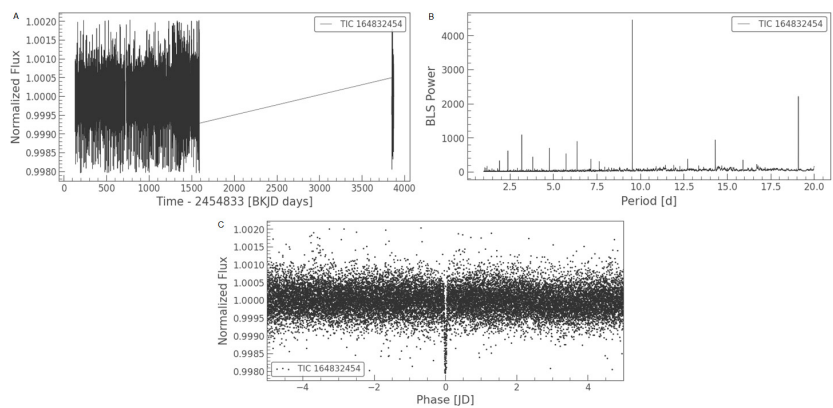

**Figure 3. Kepler-504 b flux, orbital period, and transit flux.** The sharp drop in observed flux of Kepler-504 shows how close the exoplanet Kepler-504 b is; its orbital period is only 9.5 days. **(A)** Graph of entire data of brightness (flux) of the star Kepler-504 over its whole observed time period. The gap in the time series data is due to two separate data sets. The first is from the Kepler Space Telescope, while the second data set is from TESS. **(B)** Orbital period of the planet using the BLS method. **(C)** The Transit Flux graph gives us the minimum flux value of planet Kepler-504 b, which helps us calculate its planetary properties and habitability.

We chose the host stars as G-type and M-type since most of the discovered habitable exoplanets revolve around those star types (16). The available data for exoplanets suggests that the habitable exoplanets form more around those star types and have higher chances of supporting life (11, 16). We focused on the G-type and M-type stars along with orbital radius similar to Earth's (since Earth also has a G-type host star) also because they are brighter, bigger, easy to detect, and are hence more numerous (16).

Our analysis used Python as the core program for calculations. Therefore, it might include limitations in rounding-off floating point numbers. Besides, the code returns values only up to certain decimal points (seven decimal points for $F_{trans}$ value, and 10 decimal points for planetary parameters). However, this analysis is further needed to be extended to larger datasets of exoplanets. Due to computing power limitations, we could not extend our analysis to a larger volume of datasets for multiple planetary systems and analyze all of them. Instead, we had to run the code each time for a system. With the help of parallel-computing and Machine Learning code, we will be able to analyze our hypothesis for the entire existing dataset of all exoplanets (17).

We can also assess the probability of life in these exoplanets using the PHI (8). This index is calculated using various factors such as solid substrate and atmosphere (8). The advantage of PHI is that it will help us understand whether life in some form exists if it is not water-based (7). While Kepler-504 b and Kepler-315 c offer exciting opportunities for follow up atmospheric study and characterization for the James Webb Space Telescopes, which can probe atmospheric signatures of exoplanets, PHI calculation of the two exoplanets will further constrain the habitability parameters (5, 7, 8).

**MATERIALS AND METHODS**
We used data that has been collected by Kepler and TESS missions over the years. The data is publicly available at the Mikulski Archive for Space Telescopes (MAST) portal (11). We also used the Planetary Habitability Laboratory website, maintained by the University of Puerto Rico at Arecibo (11, 16).

We used the Python programming language for all of our calculations. The transit time-series data in Flexible Image Transform System (FITS) format was imported from the MAST portal. This research used *Lightkurve*, a Python package for Kepler and TESS data analysis. Additionally, we used the *Matplotlib* package in our code to derive the parameters required for our calculations. We imported the planet's mass, stellar parameters, and cross-checked the calculated planetary physical properties from databases and catalogues such as Exoplanet Transit Database, Exoplanet. eu, and NASA's Exoplanet Archive (12).

**Planetary Analysis**
We imported the *Lightkurve* and *Matplotlib* modules of Python in our code to fetch the flux data from the MAST portal (12). The flux data is timestamped by the KST mission in units of Barycentric Kepler Julian Date (BKJD) along with the catalog number of the star (12). The catalog numbers serve as identifiers and are used to store data of the star in the MAST portal (12). KIC is the Kepler Input Catalog for stars observed by Kepler while TIC is the TESS Input Catalog number for stars observed by TESS (12).

We used the Box Least Squares (BLS) Periodogram method to calculate the orbital period of the planets (18). The BLS Periodogram is a statistical tool used for detecting transiting exoplanets in the time series data (18). It is optimized to find 'box' or 'transit-shaped' periodic signals in the time-series data.

Using data provided by the MAST Portal, we used Python to import the time series data of the host star of the exoplanet (12). The time series data of a star is its measured brightness (or flux) over a period of time. The code plotted the flux vs. time graph for the exoplanet. Then, we used the BLS (Box-Least Squares) Periodogram method to find the orbital period of the planet (18). It is an optimized method to find transit-signals in the time series data. The code calculated the orbital period of the exoplanet from the BLS vs. Period graph. We plotted the normalized flux against phase (given by Julian Day (JD)) and then extracted the minimum flux ($F_{trans}$) value. Using this $F_{trans}$ value, the planet's orbital period, and the stellar parameters, we further calculated the planet's orbital radius, radius, and surface temperature. Using those parameters, the code computed the ESI values for the exoplanets. The same analysis was done for all three exoplanets.

For the star Kepler-315, we masked the original data after finding Kepler-315 b to calculate the orbital period of the second planet Kepler-315 c using the BLS method. The masking of data is the hiding of the original data that belongs to the previous planet. Otherwise, the orbital period calculation for the next planet in that host star system cannot be done. For Kepler-504 b, all the data points from KST and TESS were normalized with the help of the *Lightkurve* package.

We did not use the radial velocity data for calculating the planet's mass. However, we took the value of the exoplanet mass from the NASA Exoplanet Catalog database and used that as an input in our Python code (12). We then computed the mass of the planet.

### Transit Photometry

The data of the star's flux (or brightness) is measured over a period of time. We looked for periodic dips in the flux, which hints at a planetary object transiting around its host star in our line of sight (3, 4, 8). In this paper, we have taken the transit time-series as the primary data for calculating those parameters.

Using this method, our Python code imported the time-series data through the *Lightkurve* module. Then, it calculated the planet's orbital period (P). From the graphs, the code extracted the $F_{trans}$. After that, it calculated the planet's orbital radius (a), planetary radius ($R_p$), and surface temperature ($T_p$) using **Equations 1**, **2**, and **3**:

$$a = \sqrt[3]{G.M_{Star}.(P/(4\pi^2)} \quad \text{(Equation 1)}$$

$$R_p = R_{Star}\sqrt{1 - F_{trans}} \quad \text{(Equation 2)}$$

$$T_p = \sqrt[4]{\frac{T_{Star}}{\sqrt{2}}[(1-\alpha)\frac{R_{Star}^2}{a^2}]} \quad \text{(Equation 3)}$$

$G$ is the Universal Gravitational Constant, where G = 6.673 x10$^{-11}$ m$^3$ kg$^{-1}$ s$^{-2}$. The variables $M_{Star}$, $R_{Star}$, and $T_{Star}$ are the host star's mass, radius, and temperature, respectively. α is the albedo and (1-α) is Coefficient of absorption. Since measuring albedo involves huge uncertainties with transit data, we have assumed α = 0.07 (according to NASA Exoplanet Archive) for all of our host stars.

### Radial Velocity

The radial velocity uses the principle of the Doppler Effect. The shifts in the observed high-resolution spectrograph of the star-planet revolution are used to calculate the planet's mass. Determining the inclination of the planet's orbit with respect to its host star, the planet's mass can be determined using Newton's Law (6). The planet's mass can also be calculated using **Equation 4**:

$$M_p = v_{max}\sqrt{\frac{aM_{Star}}{G}} \quad \text{(Equation 4)}$$

The variable $v_{max}$ is the velocity semi-amplitude obtained from plotting the velocity versus time graph. We have directly taken the value of mass of the exoplanet from the NASA Exoplanet Catalog database (12). Using that as an input in our Python code, we have used it for calculating the planetary parameters needed for our ESI calculations.

### Calculating ESI values

The ESI of a parameter $x$ ($ESI_x$) of an exoplanet was calculated using **Equation 5**:

$$ESI_x = (1 - \left|\frac{x-x_0}{x+x_0}\right|)^{w_x} \quad \text{(Equation 5)}$$

For calculating ESI values, we used the Earth's weight exponent values ($w_x$) for radius, density, escape velocity and surface temperature. These values were taken from the PHL database.

The $ESI_i$ measures the rocky interior of the planet compared to Earth (**Equation 6**) (3, 8). $ESI_i$ is calculated by taking the square root of the product of the ESI of radius ($ESI_r$) and ESI of density ($ESI_{rho}$) of the exoplanet:

$$ESI_i = \sqrt{ESI_r . ESI_{rho}} \quad \text{(Equation 6)}$$

Similarly, the $ESI_s$ measures its capacity to sustain a surface like that of Earth (**Equation 7**) (3, 8). $ESI_s$ is calculated by taking the square root of the product of ESI of escape velocity ($ESI_v$) and ESI of surface temperature ($ESI_t$) of the exoplanet:

$$ESI_s = \sqrt{ESI_v . ESI_t} \quad \text{(Equation 7)}$$

$ESI_G$ is a measure of the physical similarity of any planetary body with Earth (**Equation 8**). $ESI_G$ is calculated by taking the square root of the product of **Equation 6** and **Equation 7** of the exoplanet:

$$ESI_G = \sqrt{ESI_i . ESI_s} \quad \text{(Equation 8)}$$

**Equation 8** gives us the probability of whether the planet is habitable in terms of Earth life. A value above 60% is considered habitable (16). A value above 80% shows very high probability (16).

### ACKNOWLEDGEMENTS
S.B. acknowledges Research Discovery (RD), which is an intensive research program designed to engage motivated high school students to develop an original research project under the guidance of world-class research mentors. G.S. is grateful to her institution for the encouragement and support.

**Received:** January 3, 2022
**Accepted:** August 23, 2022
**Published:** November 07, 2022